\documentclass{ecai} 

\usepackage{latexsym}
\usepackage{amssymb}
\usepackage{amsmath}
\usepackage{amsthm}
\usepackage{booktabs}
\usepackage{enumitem}
\usepackage{graphicx}
\usepackage{color}

\usepackage{multirow}
\usepackage{makecell}
\usepackage{caption}

\usepackage{comment} 
\usepackage{listings} 
\usepackage{xcolor}   
\usepackage{tcolorbox}
\tcbuselibrary{listings, skins}

\definecolor{darkred}{rgb}{0.6,0,0} 
\definecolor{darkgreen}{rgb}{0,0.5,0} 
\definecolor{darkblue}{rgb}{0,0,0.5} 
\definecolor{deeppurple}{rgb}{0.5, 0, 0.5}
\lstdefinestyle{diffstyle}{
    basicstyle=\ttfamily\scriptsize,
    breaklines=true,
    showstringspaces=false,
    commentstyle=\color{darkred},
    morecomment=[l][\color{darkblue}]{diff},
    morecomment=[l][\color{darkgreen}]{+    },
    morecomment=[l][\color{darkred}]{-    },
    morecomment=[l][\color{deeppurple}]{@@},
    frame=tb, 
    numbers=left, 
    numberstyle=\small\color{black}, 
    captionpos=b, 
    moredelim=**[is][\bfseries]{|}{|}, 
}

\newcommand{\BibTeX}{B\kern-.05em{\sc i\kern-.025em b}\kern-.08em\TeX}

\begin{document}


\begin{frontmatter}




\title{CLNX: Bridging Code and Natural Language for C/C++ Vulnerability-Contributing Commits Identification}


\author{
    Zeqing Qin$^{\dagger \ddagger}$, Yiwei Wu$^{\dagger}$, Lansheng Han$^{\ast \dagger \ddagger}$  \\
    \small
    $^{\dagger}$School of Cyber Science and Engineering, Huazhong University of Science and Technology, Wuhan, China \\
    $^{\ddagger}$Hubei Key Laboratory of Distributed System Security, Hubei Engineering Research Center on Big Data Security \\
    \textnormal{zeqing@hust.edu.cn, cnwyw77777@gmail.com, hanlansheng@hust.edu.cn ($^{\ast}$Corresponding author)}
}


\begin{abstract}
Large Language Models (LLMs) have shown great promise in vulnerability identification. As C/C++ comprises half of the Open-Source Software (OSS) vulnerabilities over the past decade and updates in OSS mainly occur through commits, enhancing LLMs' ability to identify C/C++ Vulnerability-Contributing Commits (VCCs) is essential. However, current studies primarily focus on further pre-training LLMs on massive code datasets, which is resource-intensive and poses efficiency challenges.
In this paper, we enhance the ability of BERT-based LLMs to identify C/C++ VCCs in a lightweight manner. We propose CodeLinguaNexus (CLNX) as a bridge facilitating communication between C/C++ programs and LLMs. Based on commits, CLNX efficiently converts the source code into a more natural representation while preserving key details. Specifically, CLNX first applies structure-level naturalization to decompose complex programs, followed by token-level naturalization to interpret complex symbols. We evaluate CLNX on public datasets of 25,872 C/C++ functions with their commits. The results show that CLNX significantly enhances the performance of LLMs on identifying C/C++ VCCs. Moreover, CLNX-equipped CodeBERT achieves new state-of-the-art and identifies 38 OSS vulnerabilities in the real world. 
\end{abstract}

\end{frontmatter}


\section{Introduction}
In recent years, with the rapid growth of open-source software (OSS) applications, the number of OSS vulnerabilities has also been increasing rapidly. According to the data from the 2023 OSSRA report \citep{2023report}, in the 1,703 codebases analyzed by the Black Duck audit team, 84\% of the codebases contained at least one known open-source vulnerability, and 48\% contained high-risk vulnerabilities. Moreover, 52.13\% of reported vulnerabilities in OSS are written in C/C++ \cite{wang2023graphspd} over the past decade. As patch commit is the primary way to update code in OSS, identifying Vulnerability-Contributing Commits (VCCs) can prevent new vulnerabilities from being introduced into OSS to a large extent \citep{meneely2013patch}.

Large Language Models (LLMs), particularly those based on the BERT \citep{devlin2018bert} architecture, have demonstrated their potential to identify vulnerabilities by effectively learning code dependencies and contextual nuances \citep{xu2022systematic}. This efficacy is attributed to their bidirectional encoder architecture, which enables the models to simultaneously consider the semantics of context both preceding and following a given segment of code. However, as these models are trained initially on natural language, there is significant room for improvement in code comprehension. Current research primarily focuses on further pre-training LLMs on extensive code datasets to address this \citep{xu2022systematic}. For example, CodeBERT \citep{feng2020codebert} has been pre-trained on six programming languages: Python, Java, JavaScript, PHP, Ruby, and Go. Nonetheless, it exhibits suboptimal performance on C/C++ due to the absence of specific pre-training for these languages. More importantly, the improvements
remain marginal even after extensive further pre-training. For instance, ContraBERT \citep{liu2023contrabert}, which has undergone further pre-training based on CodeBERT, achieves only minor percentage-point improvements (a rise of 1.24\% in accuracy) in identifying C/C++ vulnerabilities while consuming significant GPU resources. It indicates that further pre-training is inefficient and occasionally ineffective \citep{liu2023pre}.

Specifically, we address the major challenge in our paper.

\begin{itemize}
\item How to enhance the effectiveness of LLMs on identifying C/C++ VCCs while ensuring a lightweight implementation?
\end{itemize}

To address this challenge, we introduce CodeLinguaNexus (CLNX), a middleware designed to translate original C/C++ code into a format that enhances compatibility with LLMs. To do so, we first perform the structure-level naturalization. Specifically, we linearize the structure of the C/C++ source code with commit and shorten their length. Then, we perform token-level naturalization. Special C/C++ symbols that differ significantly from natural language are transformed into their natural language representations.

We implement CLNX and evaluate it on a dataset of 25, 872 C/C++ functions with corresponding commits, including 10, 894 VCCs. The result shows that CLNX significantly improves LLMs' performance on C/C++ VCCs identification. Moreover, equipped with CLNX, BERT undergoes an increase of 14.48\% in precision, surpassing other models that have been further pre-training on code. Finally, the CLNX-equipped CodeBERT achieves the best effectiveness and becomes new state-of-the-art. Lastly, CLNX-equipped CodeBERT finds 38 real-world OSS vulnerabilities by identifying vulnerability-contributing commits, demonstrating CLNX’s ability to help LLMs report vulnerabilities in the real world. 

In summary, our contributions in this paper are:
\begin{itemize}
\item We propose CLNX, a pioneering framework for improving LLMs' performance on C/C++ VCCs identification in an effective and efficient way.
\item We successfully implement a prototype of CLNX and conduct extensive experiments to evaluate its effectiveness.
\item We equip CodeBERT with CLNX to achieve the new state-of-the-art and demonstrate CLNX-equipped CodeBERT's ability to identify vulnerabilities in the real world.
\end{itemize}

\section{Preliminaries}

\subsection{Vulnerability-Contributing Commits}
In OSS development, patch commits record the differences between two versions of the source code \citep{zuo2024vulnerability}. They can be categorized into two types: vulnerable patch commits and non-vulnerable patch commits. Vulnerable patch commits refer to those that will introduce new vulnerabilities into the original code, which are also called Vulnerability-Contributing Commits (VCCs) \cite{meneely2013patch}. In this research, a patch commit is considered "vulnerable" if it introduces vulnerabilities that belong to any of the Common Weakness Enumeration (CWE), regardless of its triggering conditions \cite{wang2023graphspd}. Listing 1 shows a vulnerable patch commit with code revisions marked by plus and minus signs (\texttt{+/-}) on the left side. This commit is a configuration item change aimed at improving the path settings for mutexes in the Apache HTTP Server. However, it introduces a vulnerability related to permission bypass. Vulnerable patch commits highlight critical information about vulnerabilities. When identifying VCCs at the functional level, both the patch commit and the source code of the revised function are analyzed.

\begin{table}[t]
\begin{lstlisting}[caption={An example of Vulnerability-Contributing Commit}]
|commit  08c38d0831c46ed5b62e2f83e42a4c84e111d553|
@@ -212,7 +212,7 @@
diff --git a/server/util_mutex.c b/server/util_mutex.c
- a/server/util_mutex.c
+ b/server/util_mutex.c
@@ -120 +120 @@ AP_DECLARE(apr_status_t) 
-  *mutexfile = ap_server_root_relative(pool, file);
+  *mutexfile = ap_runtime_dir_relative(pool, file);
@@ -307 +307 @@ static const char 
-  return ap_server_root_relative(p,
+  return ap_runtime_dir_relative(p,
@@ -555 +555 @@ AP_CORE_DECLARE(void) 
-  dir = ap_server_root_relative(p, mxcfg->dir);
+  dir = ap_runtime_dir_relative(p, mxcfg->dir); 
\end{lstlisting}
\end{table}


\subsection{Pre-training and Fine-tuning}
Pre-training in this paper refers to the training phase of LLMs conducted on large-scale unlabeled datasets. LLMs can generally be divided into two categories: BERT-based and GPT-based. Since GPT-based LLMs are composed of a decoder structure and are more suitable for generative tasks \citep{hadi2023survey}, we primarily focus on the performance of BERT-based models in vulnerability identification, a code classification task\citep{xu2022systematic}. BERT-based LLMs are pre-trained on tens of millions of text data using techniques like Masked Language Modeling (MLM) and Next Sentence Prediction (NSP) \citep{xu2022systematic}. During this phase, these models capture helpful information from the data and store it in their weights. These pre-trained models are then fine-tuned on labeled data for specific downstream tasks like text classification or question answering. While pre-training requires substantial computational resources, fine-tuning is comparatively more resource-efficient \cite{radiya2020fine}.

\section{Methodology}
This section presents an overview of our approach and details each component, including structure-level and token-level naturalization.
\begin{figure*}
    \centering
    \includegraphics[width=\linewidth,scale=1.00]{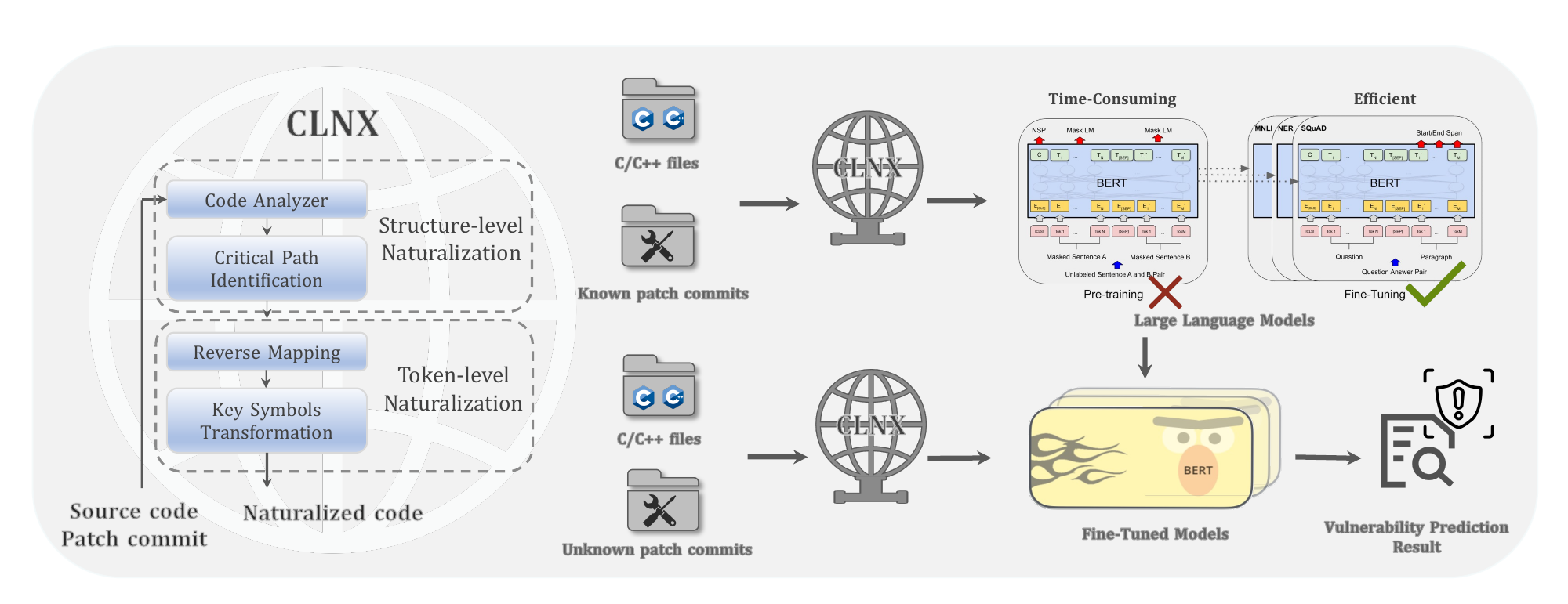}
    \caption{The Overview of CLNX}
    \label{fig:main}
    \vspace{3pt}
\end{figure*}

\subsection{Overview}

The overview of CLNX is shown in Figure \ref{fig:main}, with CLNX's internal structure displayed on its left side. CLNX is performed at the functional level. In the process of handling input source code and patch commit, CLNX initially undertakes structural-level naturalization. This stage involves employing CLNX's code analyzer to transform the source code into a graph of linear execution paths, followed by the integration of patch information to select the critical path. Subsequently, CLNX advances to token-level naturalization, which involves mapping the identified critical path to the corresponding source code and transforming key symbols into their natural language equivalents. Finally, CLNX outputs the fully naturalized version of the source code. The system workflow for deploying CLNX to enhance LLMs' performance on VCCs identification is shown on the right side of Figure \ref{fig:main}. For a given set of programs with their corresponding patch commits, the programs are naturalized by CLNX and then provided to LLMs for fine-tuning. When an unknown program with its patch commit is analyzed, CLNX transforms the program into naturalized form and then forwards the results to the fine-tuned LLMs for vulnerability identification. In the rest of Section 3, we formalize the details of each component of CLNX.

\begin{figure}[h]
    \centering
    \includegraphics[width=\linewidth]{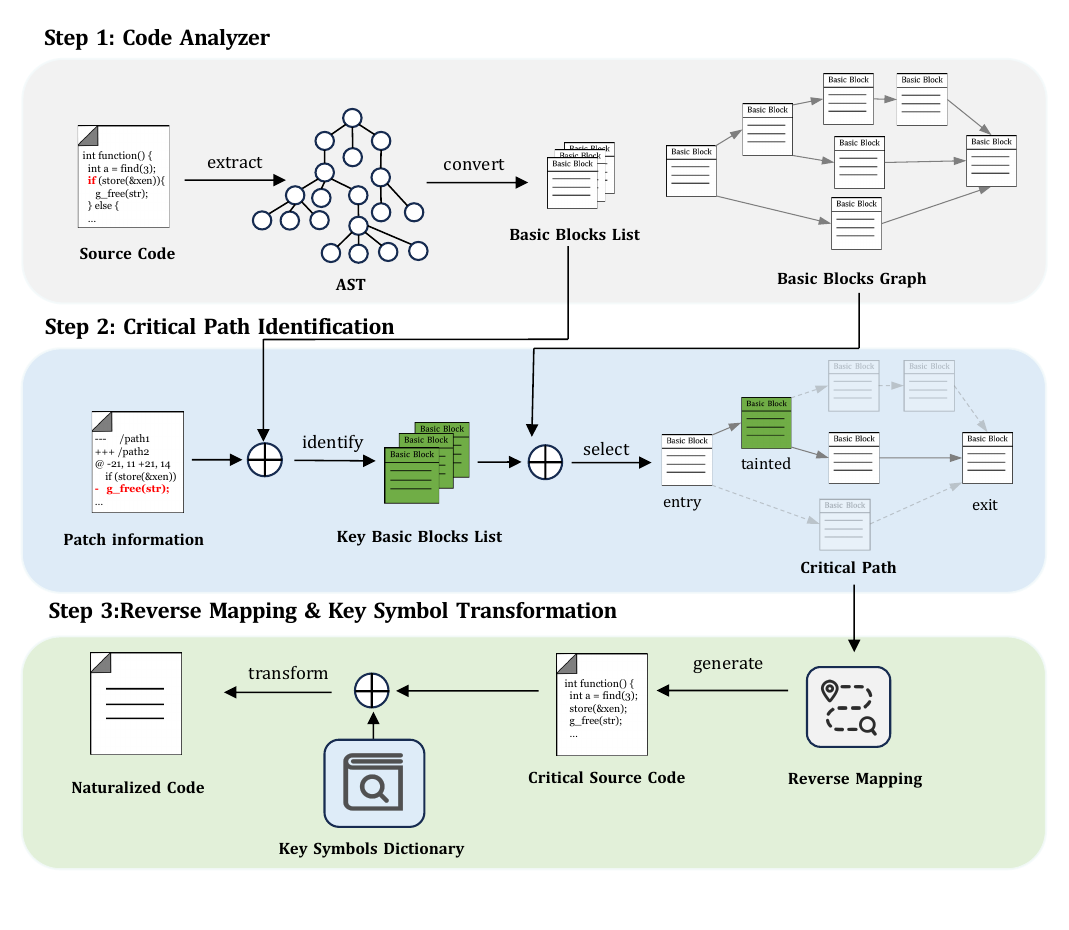}
    \caption{The Workflow of CLNX}
    \label{fig:workflow}
    \vspace{18pt}
\end{figure}

\subsection{Code Analyzer} 
C/C++ programs' complex structures and excessive length challenge LLMs in understanding them. In response to these challenges, CLNX's structure-level naturalization is designed with two primary goals: First, it linearizes complex program structures; Second, it reduces the overall program length. In particular, the code analyzer extracts linear execution paths within a program. 

In the design of CLNX's code analyzer, the concept of 'basic blocks,' as borrowed from LLVM \cite{basicblock}, plays a pivotal role. A 'basic block' is a sequence of instructions that executes sequentially, characterized by a single entry and a single exit point, devoid of any internal branching. The code analyzer transforms programs into basic blocks and generates a graph $G=(V, E)$, where each vertex in $V$ corresponds to a basic block, and each edge in $E$ represents the control flow between blocks. The graph $G$'s entrance point $v_{entry}$ corresponds to the program's entry basic block, and its exit point $v_{exit}$ corresponds to the program's final basic block. As a result, any path traversed from $v_{entry}$ to $v_{exit}$  within $G$ delineates a linear execution path of the program. In particular, when there is a loop structure, for simplicity, we directly convert the control flow to single executions and label the corresponding nodes as loop structures. It should be noted that CLNX only uses  Abstract Syntax Tree (AST) and Control Flow Graph (CFG) for code embedding. While the Program Dependence Graph (PDG), integrating both control dependency graph (CDG) and data dependency graph (DDG), is commonly used to abstractly represent source code \cite{wang2023graphspd}. We believe that complex structures risk subjectively introducing excessive irrelevant information, thereby complicating the accurate semantic representation of the code. We compare our method with complex graph-based approaches (embedding AST/CFG/DDG/CDG) in RQ2 to demonstrate CLNX's effectiveness.

The code analyzer deploys Joern to generate AST. The whole process is illustrated in Step 1 of Figure \ref{fig:workflow}. In contrast to LLVM, CLNX's code analyzer does not impose requirements on the actual compilability of the program. This attribute is particularly significant for identifying function-level vulnerabilities, especially in scenarios where the absence of relevant header files precludes successful compilation.

\subsection{Critical Path Identification} After obtaining the graph G composed of basic blocks, the focus of CLNX shifts to identifying a critical execution path within the graph that encompasses the maximum amount of vulnerability-related basic blocks. This process can be divided into two primary steps, as illustrated in Step 2 of Figure \ref{fig:workflow}; Firstly, determining the basic blocks that are directly related to a patch commit. Secondly, the critical path within G is selected, which offers the most extensive coverage of these identified basic blocks.

\subsubsection{commit-related basic blocks identification} Based on the idea of taint analysis \cite{boxler2018static}, CLNX identifies the code removed in the corresponding patch commits of a program as contamination points. CLNX also considers an extended range, which includes three lines \cite{wang2023graphspd} before and after the lines corresponding to the removed code, as the affected tainted area. This area is represented as $S = [l_{s}, l_{e}]$, where $l_{s}$ and $l_{e}$ are the start and the end line numbers of the tainted area, respectively. A basic block $BB_{i}$, covering the line number range $[b_{is}, b_{ie}]$, is regarded as commit-related, denoted  $BB_{tainted_i}$, if its range intersects with the tainted area, i.e., $\{BB_{tainted_i} \vert [l_{s},l_{e}]\cap[b_{is},b_{ie}]\not =\emptyset\}$.

\subsubsection{critical executing path selection}
Further, CLNX is going to select one critical executing path in graph G. CLNX designates the basic block corresponding to the program's entry point as the source $(BB_{source})$ and the basic block corresponding to the program's exit point as the sink $(BB_{sink})$. Based on this, the critical linear execution path $P$ in the graph structure, which originates from $BB_{source}$ and terminates at $BB_{sink}$, aims to maximize the coverage of vulnerability-related basic blocks $BB_{tainted}$. CLNX designs its $critical\_path\_selecting$ algorithm based on dynamic programming to circumvent the issue of path explosion. $critical\_path\_selecting$ algorithm selects the critical execution path in graph $G$ by satisfying three primary criteria: First, the path covers as many $BB_{tainted}$ as possible. Second, the path minimizes length.  Third, if two paths have the same length and contain the same number of $BB_{tainted}$, then select the one with the highest information entropy value. 


\subsection{Reverse Mapping}
 As a compiled programming language, C/C++ has more low-level symbols compared to natural languages. In response, CLNX's token-level naturalization is designed to translate complex symbols into their natural linguistic equivalents. Initially, CLNX undertakes the task of reverse mapping the critical path, composed of basic blocks, back to the source code. This step involves reconstructing the source code information by tracing the sequence of basic blocks within the critical path. Owing to the grand architecture of CLNX basic blocks, the implementation of reverse mapping is straightforward and efficient.

 \subsection{Key Symbols Transformation}
 CLNX designs rules to transform five types of C/C++ symbols into natural presentations. These transformations are semantic-preserving but rewrite original code symbols into artificial, natural forms. Given the source code, CLNX deploys appropriate transformations based on the symbols' type and rewrites the symbols to naturalize the source code. The procedure is illustrated in Step 3 of Figure \ref{fig:workflow}. The selection of the key symbols is motivated by the low-level characters of C/C++ \citep{lee2018reconciling}.
The examples of transformation rules are shown in Table \ref{tab:my_label_2}.

\textbf{Operator Symbols}: Operators directly influence CPU computation instructions. For instance, logical operators involve the CPU's logical instructions; bitwise operators operate directly on the bits of operands. Some operator symbols in C/C++ are high abstraction and symbolization, closer resemblance to low-level machine language, and semantic complexity. CLNX identifies key operator symbols, including Pointer Operator, Bitwise Operator, and Shift Operator.

\textbf{API Call Symbols}: API call functions involve interactions between the program and the runtime environment, forming the basis for the program's proper functioning. In C/C++, specialized API call symbols pose challenges for LLMs due to their close integration with underlying systems. CLNX identifies key API call symbols, including Memory Management API Calls, Synchronization Mechanisms API Calls, and System Calls.

\textbf{Control Flow Symbols}: Control flow symbols directly affect the execution path of a program. C/C++ owns some unique control flow symbols. CLNX identifies key control flow symbols, including Setjmp/Longjmp and Goto.

\textbf{Preprocessor Directive Symbols}: Preprocessor directives are a part of the compilation process, executing before the compiler compiles the source code. Preprocessors allow for conditional compilation of different code segments based on specific conditions. This technique, common in C/C++, addresses code compatibility issues across various platforms and compilation environments. CLNX identifies key preprocessor directive symbols, including Header Files, Macro Definitions, Conditional Compilation, and Preprocessor Logic.

\textbf{Declaration Symbols}: Declaration defines a program's data structures and memory allocation. It allows the compiler to perform type checking, prevent type errors, and optimize at the lower level. Declarations in C/C++ have distinct features, such as low-level and complex. CLNX identifies declaration symbols covering Basic Data Types, Classes, and Templates.

\begin{table}[t]
    \centering
    \caption{Key Symbols to be transformed}
    \label{tab:my_label_2}
    \fontsize{8pt}{10pt}\selectfont 
    \resizebox{\columnwidth}{!}{%
    \begin{tabular}{@{}cll@{}}
    \toprule
    Type & Example & \makecell{Natural Language Equivalents}\\
    \midrule
    \multirow{8}{*}{Operator} & *p & dereference p\\
    & \&var & obtain address of var \\
    & a | b & a Bitwise OR b \\
    & a \textasciicircum b & a Bitwise XOR b \\
    & \textasciitilde a & Bitwise NOT of a \\
    & a \& b & a Bitwise AND b\\
    & a << b & a left shift by b\\
    & a >> b & a right shift by b\\
    \midrule
    \multirow{4}{*}{API Call} & malloc(size) & allocate memory of size\\
    & free(ptr) & deallocate memory of ptr\\
    & pthread\_create(t) & create new thread t\\
    & write(data) & write data to file descriptor\\
    \midrule
    \multirow{2}{*}{\makecell{Control \\Structure}} & goto label & jump to the statement label \\
    & setjmp(env) & save the current environment env \\
    \midrule
    \multirow{2}{*}{\makecell{Preprocessor\\ Directive}} & \#include <h> & include header file <h>\\
    & \#elif condition & else if condition\\
    \midrule
    \multirow{3}{*}{\makecell{Declaration}} & struct P{} & declare a structure P{}\\
    & template<T> & template class definition <T>\\
    & volatile int s & declare volatile variable s\\
    \bottomrule
    \end{tabular}%
    }
\end{table}

\section{Experimental Setup}

Our evaluation is designed to answer the following research questions:
\begin{itemize}[leftmargin=*]
\item RQ1: How does CLNX enhance LLMs for the C/C++ VCCs identification task?
\item RQ2: How does the performance of CLNX-equipped LLMs compare to other vulnerability identification-related  methods?
\item RQ3: How does CLNX-equipped LLM perform in identifying real-world OSS vulnerabilities that are contributed through commits?
\end{itemize}

\subsection{Evaluation Task}

The evaluation task of our paper is Vulnerable-Contributing Commits (VCCs) identification, where the input is the source code and the corresponding patch commit, and the output is a label denoting whether the commit will introduce vulnerabilities into the original code or not. 

\subsection{Datasets}

To evaluate our research questions using real-world data, we construct our experimental datasets based on the publicly released version of the Devign dataset \cite{zhou2019devign}, which includes both vulnerable and non-vulnerable functions, along with their associated commit IDs, from two major open-source C/C++ projects: FFmpeg and Qemu. 

\subsection{Evaluation Metrics}

In our experiments, different metrics are used to evaluate downstream tasks. We follow the metrics that CodeXGLUE \cite{lu2021codexglue} used for evaluation, and the details are listed below:

\begin{itemize}[leftmargin=*]
\item Prec: Precision measures the proportion of correct positive identifications made by the model compared to the total predicted positives.
\item Acc: Accuracy defines the ratio of correct predictions (i.e., the exact match) in the test set.
\item Recall: This metric concentrates on the model's ability to correctly identify all genuine positive instances. It calculates the proportion of true positives accurately detected by the model out of the total positives.
\item F1: This metric is the harmonic mean of precision and recall, balancing these two metrics. It is advantageous when class distribution is imbalanced.
\end{itemize}

\subsection{Baselines}

We consider both BERT-based and GPT-based LLMs for evaluation, and we mainly focus on CLNX's effectiveness in improving BERT-based LLMs. This is because BERT-based LLMs' proficiency in comprehensively understanding vulnerability through their bidirectional encoder structure. For comparison, we include vulnerable patch commit identification methods, deep learning vulnerability identification methods, and traditional vulnerability identification tool. The details of the baselines are listed in Table \ref{table_baseline}. It should be noted that, when LLMs without the equippment of CLNX, deep learning vulnerability identification methods, and traditional vulnerability identification tools are applied to VCCs identification, they directly take the revised code after patch committing as input to identify whether it is vulnerable. We make sure that the vulnerabilities of these revised code are only contributed by corresponding patch commits.

\begin{table}[t]
    \centering
    \fontsize{8pt}{10pt}\selectfont
    \caption{Details of Baselines}
    \label{table_baseline}
    \resizebox{\columnwidth}{!}{%
    \begin{tabular}{ccl}
    \toprule
    Type & Specific & Baseline  \\
    \midrule
    \multirow{7}{*}{\makecell{LLM}}   & \multirow{5}{*}{\makecell{BERT-based}}  & BERT \citep{devlin2018bert}\\
    &                                                                           & DistilBERT \citep{sanh2019distilbert}  \\
    &                                                                           & RoBERTa \citep{liu2019roberta} \\
    &                                                                           & ContraBERT \cite{liu2023contrabert} \\
    &                                                                           & CodeBERT \cite{feng2020codebert} \\
    & \multirow{2}{*}{\makecell{GPT-based}}                                     & GPT-3.5 Turbo \\
    &                                                                           & GPT-4.0 \cite{achiam2023gpt} \\
    \midrule
    \multirow{2}{*}{\makecell{Vulnerable patch commit\\ identification method}} & Token-based & VulFixMiner \citep{zhou2021finding} \\
                                                                                & Graph-based & GraphSPD \citep{wang2023graphspd} \\
    \midrule
    \multirow{5}{*}{\makecell{Deep learning vulnerability\\ identification method}} &  \multirow{3}{*}{\makecell{Token-based}}                                                    & Russel \citep{russell2018automated} \\
     &                                                                          & VulDeePecker \citep{li2018vuldeepecker} \\
     &                                                                          & SySeVR \citep{li2021sysevr} \\
     &  \multirow{2}{*}{Graph-based}                                            & Devign \citep{zhou2019devign} \\
     &                                                                          & REVEAL \citep{chakraborty2021deep}  \\
     \midrule
     Traditional tool                 & Static Analysis                         &Cppcheck \citep{pereira2020use} \\
    \bottomrule
    \end{tabular}%
    }
\end{table}

\subsection{Experimental Settings}

In our evaluation tasks, we utilize the established configuration parameters for LLMs following the standardized settings provided by CodeXGLUE \cite{lu2021codexglue}. All the compared methods are re-implemented to adhere to the default specifications outlined in their foundational papers. Our implementation of CLNX utilizes Joern v2.0.120 and Scala v3.3.1. All operations of CLNX, including the code analyzer, critical path selection, and key symbol transformation, are executed on an Intel Xeon(R) Gold 6326 CPU @ 2.90GHz. We perform LLMs fine-tuning on a dedicated machine with an NVIDIA Tesla A100 GPU featuring 64GB of memory. The fine-tuning parameters and process are strictly in accordance with the defect-detection subject of CodeXGLUE \cite{lu2021codexglue}, where the epoch is 10, the block size is 400, the train batch size is 32, the eval batch size is 64, and the learning rate is 2e-5.

\section{Experimental Result}\label{heads}

\subsection{RQ1: Effectiveness}
We conduct extensive experiments and an ablation study to assess the effectiveness of CLNX's two sequential naturalization phases in enhancing LLMs' ability to identify C/C++ VCCs. It should be noted that RoBERTa, ContraBERT, and CodeBERT have undergone further pre-training with programming data. The results, including precision, accuracy, recall, and F1 score, are presented in Table \ref{tab:my_label_6}.  'with CLNX\_S' denotes models equipped only with CLNX's structure-level naturalization, while 'with CLNX' signifies models that completed both naturalization phases.

\begin{table}[t]
    \centering
    \caption{Results of LLMs on C/C++ VCCs identification}
    \label{tab:my_label_6}
 
    \fontsize{8pt}{10pt}\selectfont 
    \resizebox{\columnwidth}{!}{%
    \begin{tabular}{@{}ccccc@{}}
    \toprule
    Technique & Prec & Acc & Recall & F1 \\
    \midrule
    GPT-3.5 Turbo & 16.78\% & 31.88\% & 11.05\% & 34.84\% \\
    GPT-4.0 & 37.08\% & 42.68\% & 33.16\% & 42.05\% \\
    \midrule
    BERT & 58.60\% & 59.85\% & 48.94\% & 54.66\% \\
    BERT with CLNX\_s & 70.33\%$\uparrow$ & 62.53\%$\uparrow$ & 46.52\% & 55.99\%$\uparrow$ \\
    BERT with CLNX & 73.08\%$\uparrow$ & 63.19\%$\uparrow$ & 49.91\%$\uparrow$ & 59.98\%$\uparrow$ \\
    \midrule
    DistilBERT & 63.94\% & 61.47\% & 46.56\% & 53.88\% \\
    RoBERTa & 65.85\% & 61.21\% & 47.64\% & 55.28\% \\
    ContraBERT & 64.78\% & 63.89\% & 48.92\% & 55.74\% \\
    CodeBERT & 66.89\% & 62.18\% & 45.16\% & 53.91\% \\
    CodeBERT with CLNX\_s & 71.66\%$\uparrow$ & 63.97\% $\uparrow$& 43.47\% & 53.95\% $\uparrow$\\
    CodeBERT with CLNX & \color{red}\textbf{75.16\%}$\uparrow$ & \textbf{\color{red}65.47\%}$\uparrow$ & \textbf{\color{red}51.83\%}$\uparrow$ & \textbf{\color{red}60.64\%} $\uparrow$\\
    \bottomrule
    \end{tabular}%
    }
\end{table}

From Table \ref{tab:my_label_6}, we can see that GPT-based models do not perform well on this task, so we mainly focus on BERT-based LLMs. There are significant improvements in C/C++ VCCs identification for BERT and CodeBERT after the sequential deployment of CLNX's two-phase naturalization. Specifically, BERT's precision improved by 14.48\%, and CodeBERT's by 8.27\%, with CLNX-equipped CodeBERT outperforming all LLMs across all metrics, highlighting CLNX's impact. Although BERT’s initial precision (58.60\%) is relative low compared to CodeBERT (66.89\%), BERT with only CLNX's structure-level naturalization achieves a precision result of 70.33\%. It surpasses all the models that have been further pre-trained on program data, including CodeBERT and RoBERTa. These results directly validate that CLNX yields a better effect compared to pre-training strategies. We attribute this improvement to CLNX's effectiveness in simplifying complex structures and emphasizing critical vulnerability information. However, we notice that although accuracy and precision values are improved for both BERT and CodeBERT after CLNX's structure-level naturalization, the recall values decreased by 2.42\% and 1.69\%, respectively. These results suggest that the models miss some vulnerabilities. We believe this phenomenon is caused by CLNX's mission to reduce the source code length. In CLNX's structure-level naturalization stage, it excessively prioritizes program length reduction when dealing with multiple paths with consistent coverage of critical nodes, which may result in the loss of certain vulnerability-related information. Yet, the complete CLNX process eventually led to the highest recall rates for both models, indicating the token-level naturalization phase's effectiveness in enhancing the understanding of retained information.\\
\indent\textbf{Answer to RQ1}: Both the structure-level and token-level naturalization phases play crucial roles in CLNX's effectiveness. CLNX enhances LLMs' performance in C/C++ VCCs identification significantly.

\subsection{RQ2: Comparision}

To further evaluate the performance of CLNX-equipped LLM, we compare it with popular deep learning vulnerability identification methods, traditional tools, and vulnerable commit identification methods. We use CodeBERT as the base model for this comparison. The results are presented in Table {\ref{tab:my_label_7}}.

As shown in Table {\ref{tab:my_label_7}}, CLNX-equipped CodeBERT significantly outperforms all the compared methods in precision (improve 10.59\%), accuracy, and F1 score. This result suggests that CLNX-equipped LLM achieves a new state-of-the-art in this task. Notably, CLNX-equipped CodeBERT excels over three graph-based methods (GraphSPD, Devign, REVEAL) that use complex code embedding methods. This success can be attributed to two factors: first, the BERT-based LLMs can perform comprehensive code analysis by considering surrounding elements like variables and functions. Second, CLNX's simple code embedding method allows LLMs to emphasize key semantic information and operate more efficiently, addressing the redundancy issue often found in graph-based models.

\begin{table}[t]
    \centering
    {\caption{Results of comparative analysis}\label{tab:my_label_7}}
    \fontsize{8pt}{10pt}\selectfont 
    \resizebox{\columnwidth}{!}{%
    \begin{tabular}{@{}ccccc@{}}
    \toprule
    Technique & Prec & Acc & Recall & F1 \\
    \midrule
    Cppcheck       & 37.02\% & 50.65\% & 17.13\% & 23.43\% \\
    \midrule
    GraphSPD & 64.57\% & 62.65\% & 40.75\% & 50.12\% \\
    VulFixMiner & 50.35\% & 53.61\% & 11.72\% & 19\% \\
    \midrule
    Russell et al. & 53.02\% & 57.93\% & 39.67\% & 45.38\% \\
    VulDeePecker & 48.42\% & 53.55\% & 26.40\% & 34.17\% \\
    SySeVR & 48.52\% & 52.67\% & 64.67\% & 55.44\% \\
    REVEAL & 56.95\% & 62.43\% & 67.80\% & 59.76\% \\
    Devign & 53.62\% & 58.62\% & 61.44\% & 57.26\% \\
    CodeBERT with CLNX & \textbf{\color{red}75.16\%} & \textbf{\color{red}65.47\%} & 51.83\% & \textbf{\color{red}60.64\%} \\
    \bottomrule
    \end{tabular}%
    }
\end{table}

\begin{table*}[t]
  {\caption{Results of finding Real-World Vulnerabilities}\label{table: Real World vulnerabilities}}
  \fontsize{8pt}{10pt}\selectfont 
  \begin{minipage}{0.48\textwidth}    
  \centering
  \setlength{\tabcolsep}{3pt} 
    \resizebox{\columnwidth}{!}{%
  \begin{tabular}{@{}ccl@{}} 
    \toprule
    Target product & CWE type  & Vulnerable file in the product\\
    \midrule   
civetweb  & CWE-125 & src/civetweb.c \\
ImageMagick  & CWE-20 & coders/cals.c \\
ImageMagick  & CWE-476 & coders/xcf.c \\
illumos-gate  & CWE-476 & fs/smbsrv/smb2\_flush.c \\
jasper  & CWE-476 & src/libjasper/jp2/jp2\_cod.c \\
json-c  & CWE-310 & json\_tokener.c \\
krb5  & CWE-189 & libkdb\_ldap/ldap\_principal2.c \\
leptonica  & CWE-119 & prog/htmlviewer.c \\
libgd  & CWE-119 & src/gd.c \\
libtiff  & CWE-119 & libtiff/tif\_next.c \\
libxkbcommon  & CWE-416 & src/xkbcomp/ast-build.c \\
linux  & CWE-119 & fs/ioctl.c \\
linux  & CWE-415 & net/ipv4/inet\_connection\_sock.c \\
linux  & CWE-200 & net/bluetooth/rfcomm/sock.c \\
linux  & CWE-200 & fs/udf/namei.c \\
linux  & CWE-200 & drivers/media/media-device.c \\
linux  & CWE-200 & net/rds/recv.c \\
linux  & CWE-20 & net/bluetooth/bnep/sock.c \\
linux  & CWE-20 & net/bridge/netfilter/ebtables.c \\
    \bottomrule
    \end{tabular}%
    }
    \end{minipage}    
    \hfill
    \begin{minipage}{0.48\textwidth}
  \centering
    \resizebox{\columnwidth}{!}{%
  \begin{tabular}{@{}ccl@{}} 
    \toprule
    Target product  & CWE type & Vulnerable file in the product\\
    \midrule
linux  & CWE-476 & crypto/rng.c \\
linux  & CWE-362 & security/keys/keyctl.c \\
linux  & CWE-362 & fs/dcache.c \\
linux  & CWE-399 & fs/ext4/super.c \\
linux  & CWE-416 & drivers/usb/serial/console.c \\
linux  & CWE-399 & asm/arch\_timer.h \\
mapserver  & CWE-119 & mapogcfilter.c \\
media-tree  & CWE-264 & mm/mremap.c \\
miniupnp  & CWE-476 & miniupnpd/upnpsoap.c \\
openjpeg  & CWE-119 & src/lib/openjp2/j2k.c \\
php-src  & CWE-476 & ext/wddx/wddx.c \\
php-src  & CWE-119 &  ext/standard/dns.c            \\
radare2  & CWE-119 & libr/bin/format/elf/elf.c \\
radare2  & CWE-125 & libr/bin/file.c \\
radare2  & CWE-125 & libr/asm/p/asm\_x86\_nz.c \\
radare2  & CWE-416 & libr/core/cbin.c \\
tcpdump  & CWE-119 & print-vqp.c \\
util-linux  & CWE-362 & login-utils/su-common.c \\
WavPack  & CWE-125 & src/open\_utils.c \\

    \bottomrule
    \end{tabular}%
    }
    \end{minipage}
\end{table*}

\textbf{Answer to RQ2}: With the help of CLNX, LLM achieves a new state-of-the-art in C/C++ VCCs identification. The simple and lightweight code embedding approach of CLNX enables the LLM to capture key semantic information effectively.

\subsection{RQ3: Real World Vulnerabilities}
Finally, to evaluate the performance of CLNX-equipped LLM on real-world vulnerabilities, we conduct an evaluation using the fine-tuned CLNX-equipped CodeBERT to scan the repositories of 35 C/C++ open-source projects. Finally, CLNX successfully detects 38 vulnerabilities in those repositories. The results are shown in Table \ref{table: Real World vulnerabilities}, where vulnerabilities cover types of Improper Permission Assignment for Critical Resource (CWE-264), Cryptographic Issues (CWE-310), Information Disclosure (CWE-200), Null Pointer Dereference (CWE-476), Out-of-Bounds Read (CWE-125), Resource Management Errors (CWE-399), Buffer Error (CWE-119), Race Condition (CWE-362), Improper Input Validation (CWE-20), Use After Free (CWE-416), Numeric Errors (CWE-189), and Double Free (CWE-415).

The results indicate that CLNX-equipped CodeBERT can identify vulnerabilities of real-world C/C++ open-source projects introduced by commits. Furthermore, we observe that the model is proficient at identifying specific types of vulnerabilities, which can be ascribed to the CLNX's capability to distill critical information from vulnerability functions, thereby aiding CodeBERT in learning the specific patterns of these vulnerabilities. For instance, the model detected six Null Pointer Dereference (CWE-476) vulnerabilities and nine Buffer Error (CWE-119) vulnerabilities, which become more apparent without extraneous information. We attribute this to CLNX's effectiveness in refining key information from vulnerability functions, thus reducing the interference of irrelevant information on LLMs. However, the model only detects one Cryptographic Issue (CWE-310). This result is because vulnerabilities of such type often involve complex processing logic and do not have relatively uniform patterns.\\
\indent\textbf{Answer to RQ3}: CLNX-equipped CodeBERT effectively finds real-world vulnerabilities in open-source C/C++ repositories, demonstrating CLNX's potential to help LLMs report 0-day C/C++ vulnerabilities in OSS.

\section{Disscussion}

This section discusses the implications, limitations, and potential threats to the validity of our work.

\subsection{Implications}

We propose a novel, cost-effective framework that enhances the effectiveness of LLMs in identifying C/C++ VCCs. The findings in our research are expected to inspire researchers to improve LLMs' ability to identify VCCs across more programming languages. CLNX offers guidelines for improving LLMs' performance on VCCs identification of specific programming languages in a lightweight manner, moving beyond the traditional reliance on extensive pre-training, which requires substantial computational resources.

 \subsection{Limitations}

The experimental results demonstrate that CLNX significantly enhances the performance of LLMs in VCCs identification. The advancement is mainly due to CLNX's effective two-stage naturalization, making the code more compatible for LLMs. However, challenges arise from a decline in the Recall score, mainly due to its structure-level naturalization, which might inadvertently omit important code information. When confronted with multiple paths having equivalent coverage of tainted basic blocks, CLNX's critical path-selecting algorithm prioritizes the shortest path for length minimization at the risk of overlooking important details. A more effective approach could involve considering data flow more substantially in the critical path selection process. However, it involves dynamic program analysis. And we will explore it in our future work.

\subsection{Threats to Validity}

\textbf{Internal Validity:} Our analysis identifies two potential threats to internal validity. Firstly, the uniform standard requirement of CLNX's code analyzer necessitates standardizing source code format before its use. Secondly, CLNX calculates path length by counting the number of basic blocks, assuming each block adds uniformly to the total length. To maintain algorithmic integrity in our critical key path selection algorithm, all edges of the input graph structure must be of equal length (by default, set to one). \\
\textbf{External Validity:} Regarding external validity, the performance of the original GPT-based LLMs is significantly lower than that of BERT-based models, so we mainly focus on how CLNX improves BERT-based LLMs' performance on C/C++ VCCs identification.

\section{Related Work}
\textbf{Large Language Models}: In recent years, there has been a notable emergence of LLMs, which are increasingly recognized as promising solutions for the field of vulnerability identification \cite{thapa2022transformer} \cite{ding2022velvet} \cite{feng2020codebert} \cite{hanif2022vulberta}. BERT \cite{devlin2018bert} is a deep bidirectional encoder based on the transformer architecture, pre-trained by Google on a vast corpus comprising millions of text passages and billions of words. BERT-based LLMs are usually pre-trained on two tasks: Masked Language Model (MLM) and Next Sentence Prediction (NSP), thus equipping them with robust semantic understanding and endowing them with substantial knowledge, making it suitable for fine-tuning on specific tasks with limited data, such as vulnerability identification \citep{xu2022systematic}. Instances of successful applications include BERT's superior detection accuracy on the SARD database compared to traditional machine learning models such as LSTM or BiLSTM. Likewise, CodeBERT \cite{feng2020codebert} and its derivatives, DistilBERT \cite{sanh2019distilbert}, RoBERTa \cite{liu2019roberta}, ContraBERT \cite{liu2023contrabert}, by further pre-training, improve the performance of LLMs on the programming language to some extent.\\
\textbf{Deep Learning Vulnerability Identification}: These methods train various deep learning models with existing datasets \cite{steenhoek2023empirical} \cite{russell2018automated} \cite{okun2013report} \cite{black2018software} \cite{booth2013national}. Subsequently, these models are deployed to identify undetected vulnerabilities. They generally fall into two primary categories: token-based methods \cite{li2018vuldeepecker} \cite{russell2018automated} \cite{li2021sysevr} and graph-based \cite{zhou2019devign} \cite{chakraborty2021deep}. Token-based approaches process the source code as sequences of tokens, leveraging models such as RNN \cite{li2021vuldeelocator} \cite{li2021sysevr} \cite{li2018vuldeepecker} \cite{zhang2019novel}, CNN \cite{russell2018automated}, and MLP \cite{coimbra2021using} for training purposes. Some strategies utilize code slices to distill pivotal information. Conversely, graph-based methods seek to encapsulate the source code's multifaceted information into graphs, then analyze using various GNN \cite{cheng2021deepwukong} \cite{cao2022mvd}. For example, the Code Property Graph (CPG) leverages information from abstract syntax trees, control flow graphs, and program dependency graphs to model the combined semantic and syntactic information of a program.\\
\textbf{Patch Commit Identification}:
In OSS, code commits serve as the core building block units of a version control system in software development \citep{zuo2024vulnerability}. The patch commit (i.e., code changes + description of changes), or patch for short, is a general concept involving modifications that are specifically focused on code updates, such as introducing new features. However, this process may introduce new vulnerabilities into the original code. To address this, a significant amount of work has focused on patch commit analysis targeting vulnerability discovery \citep{zuo2024vulnerability} \citep{meneely2013patch}. In the early stage, hand-crafted features-based methods are proposed. For example, VCCFinder \cite{perl2015vccfinder} utilized an SVM model to automatically identify commits that might introduce vulnerabilities. Wang et al. \citep{wang2019detecting} studied code diffs exclusively, employing 61 features, including 22 from previous work \citep{tian2012identifying}, to form an input vector for their machine learning model. In recent years, advancements in neural networks, particularly in natural language processing (NLP) and applied graph theory, have revolutionized this field. E-SPI \citep{wu2022enhancing}, for instance, analyzes both code diffs and commit messages by first extracting a contextual abstract syntax tree (AST) from code changes, then encoding it into paths using a BiLSTM. Commit messages are converted into graphs and processed with a graph neural network (GNN). However, the quality of commit messages can limit the usefulness of such analyses. In VulFixMiner \citep{zhou2021finding}, the authors only consider code change information. To extract semantics from the code changes, they adopt CodeBERT. It is noteworthy that VulFixMiner only investigates Python and Java projects \citep{zuo2024vulnerability}. Most recently, a detection system called GraphSPD is proposed \citep{wang2023graphspd}, which proposes a novel graph structure called PatchCPG to represent patches. Then, it applies an end-to-end deep learning model called PatchGNN to classify patch commits, and it achieves the state-of-the-art.

\section{Conclusion}

In this research, we propose CLNX, a middleware framework that naturalizes C/C++ programs to be compatible with LLMs, thereby improving their ability to identify C/C++ VCCs. Since CLNX requires no GPU resources at all, it is very efficient compared to pre-training. Extensive experiments confirm that CLNX-equipped LLMs demonstrate robust improvements in C/C++ VCCs identification, achieving new state-of-the-art. We anticipate that CLNX will allow developers to effectively improve the performance of LLMs in identifying VCCs of specific programming languages without additional pre-training.





\bibliography{main}

\end{document}